\documentclass[]{interact}

\usepackage{natbib}
\usepackage{caption}
\usepackage{xcolor}
\usepackage{enumerate}
\usepackage{epstopdf}% To incorporate .eps illustrations using PDFLaTeX, etc.
\usepackage[caption=false]{subfig}% Support for small, `sub' figures and tables
\usepackage{algpseudocode}
\bibpunct[, ]{(}{)}{;}{a}{,}{,}% Citation support using natbib.sty
\renewcommand\bibfont{\fontsize{10}{12}\selectfont}% To set the list of references in 10 point font using natbib.sty

\usepackage{lineno}
\usepackage{amsmath}

\usepackage{algorithm}
\usepackage{algpseudocode}
\usepackage{array}
\usepackage{graphicx}
\usepackage{url}
\usepackage{soul}

\theoremstyle{plain}% Theorem-like structures provided by amsthm.sty

\theoremstyle{definition}

\theoremstyle{remark}

\usepackage{geometry}
\setcitestyle{authoryear,open={(},close={)}}

\begin{document}

\articletype{Research Article}

\title{Bringing Spatial Interaction Measures into Multi-Criteria Assessment of Redistricting Plans Using Interactive Web Mapping}

% don't include in actual submission - just for our reference
%\tableofcontents

% ORCID
% Jacob Kruse: 0000-0001-7752-3190
% Song Gao: 0000-0003-4359-6302
% Yuhan Ji: 0000-0002-4426-6105

\author{\name{Jacob Kruse \textsuperscript{a}, Song Gao \textsuperscript{a*}, Yuhan Ji \textsuperscript{a}, Daniel P. Szabo \textsuperscript{b}, Kenneth R. Mayer \textsuperscript{c}} \thanks{*A preprint draft and the final version will be available on the journal of Cartography and Geographic Information Science; Corresponding Author: Song Gao, song.gao@wisc.edu}
\affil{\textsuperscript{a}Geospatial Data Science Lab, Department of Geography, University of Wisconsin-Madison \textsuperscript{b}E\" otv\"os L\'or\'and Tudom\'anyos Egyetem, Department of Operations Research, Budapest}\textsuperscript{c}Department of Political Science, University of Wisconsin-Madison}
% \thanks{Corresponding author: song.gao@wisc.edu}
\maketitle

%\newpage

% Make sure abstract is distinct from recom introduction paper and within 250 words
\begin{abstract}

Redistricting is the process by which electoral district boundaries are drawn, and a common normative assumption in this process is that districts should be drawn so as to capture coherent communities of interest (COIs). While states rely on various proxies for community illustration, such as compactness metrics and municipal split counts, to guide redistricting, recent legal challenges and scholarly works have shown the failings of such proxy measures and the difficulty of balancing multiple criteria in district plan creation. To address these issues, we propose the use of spatial interaction communities to directly quantify the degree to which districts capture the underlying COIs. Using large-scale human mobility flow data, we condense spatial interaction community capture for a set of districts into a single number, the interaction ratio (IR), which can be used for redistricting plan evaluation. To compare the IR to traditional redistricting criteria (compactness and fairness), and to explore the range of IR values found in valid districting plans, we employ a Markov chain-based regionalization algorithm (ReCom) to produce ensembles of valid plans, and calculate the degree to which they capture spatial interaction communities. Furthermore, we propose two methods for biasing the ReCom algorithm towards different IR values. Using the ensembles produced from these three methods, we perform a multi-criteria assessment of the space of valid maps, and present the results in an interactive web map. The experiments on Wisconsin congressional districting plans demonstrate the effectiveness of our methods for biasing sampling towards higher or lower IR values. Furthermore, the analysis of the districts produced with these methods suggests that districts with higher IR and compactness values tend to produce district plans that are more proportional with regards to seats allocated to each of the two major parties.

\end{abstract}


\newpage

%\linenumbers
% Focus on methods and results first!

\section{Introduction}

% \subsection{Political Redistricting}
In the United States, voters within a given electoral district elect a representative to protect and advance their interests in the government. Political redistricting is the process by which a set of geographic units (e.g., census block groups) are partitioned into electoral districts of equal population, producing a map with district boundaries~\citep{williams1995political}. Though most redistricting rules in the United States were written before the invention of computers, maps are now made and contested with the aid of computers \citep{altman2005crayons, turner1999us}, and regionalization algorithms \citep{deford2019recombination, aydin2018skater, guo2008regionalization}. Advances in computationally-driven redistricting have, however, exposed the flaws of various redistricting criteria, as well as the conflicts that arise when trying to accommodate multiple criteria. As a result, there has been a vigorous political, academic, and legal debate about the means and ends of political redistricting \citep{nagle2019criteria,winburn2008realities, webster2013reflections}.

% spatial interactions review
A central premise in the debate on redistricting is that districts should be designed so as to capture coherent underlying communities, commonly referred to as communities of interest (COIs). However, there is no agreed upon method for defining such communities, and, consequently, a variety of proxy measurements are used instead. Borrowing from the spatial analysis and economic literature, we propose that spatial interactions be used to quantify COIs for the purposes of redistricting, as COIs defined in this way would be directly based on the spatial distribution of human activities, rather than proxies for it.

With a method to quantitatively define COIs captured in redistricting plans, it is also possible to evaluate how COI for a given plan varies with other common redistricting criteria, such as compactness and efficiency gap (see section \ref{sec:metrics}). To perform a multi-criteria analysis of redistricting plans, we employ and modify a state-of-the-art algorithm, the Recombination algorithm \citep{deford2019recombination} to produce ensembles of valid district plans. Finally, as redistricting is a political process that garners attention from the public, legal scholars, and researchers alike, it is important to engage and educate these various groups regarding any proposed redistricting methodologies. Accordingly, we present a public-facing web map which can be used to understand our multi-criteria assessment on various redistricting plans, in particular regarding the trade-offs in other redistricting metric values when trying to optimize for a particular metric. 

% \subsection{Research Contributions}

Bringing together these different research threads, the major contributions of this work are as follows:
\begin{enumerate}
    \item Theoretical: the novel use of spatial interactions as the basis of COIs in redistricting, quantified in a global way with the interaction ratio (IR).
    \item Technical: two strategies for producing district plan ensembles in which higher or lower spatial interaction values are preferentially sampled with the Recombination (ReCom) sampling algorithm.
    \item Educational: an interactive mapping website for exploring trade-offs between various redistricting criteria, containing functionality not present in existing redistricting websites.
\end{enumerate}

The remainder of this paper is organized as follows. In section 2 we present a literature review of measurements for COIs, spatial interactions in network-based community detection, redistricting algorithms and district evaluation methods. In section 3, we describe the human mobility flow dataset, and how it is used to calculate the spatial interaction ratio in this study. Next, we describe how to bias the ReCom algorithm sample distribution in the space of redistricting plans towards districts with higher or lower interaction ratios. In section 4, we present experiments used to generate redistricting plans for Wisconsin congressional
district maps, which are then used to verify the effectiveness of the proposed sampling strategies and compare the interaction ratio with other redistricting metrics, as well as a usability study on the developed web map. In section 5 we discuss limitations of this work. Finally, in section 6, we conclude this work and suggest areas for future research.

\section{Related Works}\label{sec:literaturereview}
\subsection{Measurements for Community of Interest Quantification}
A common normative assumption in redistricting is that districts should be drawn so as to capture coherent underlying communities, commonly referred to as communities of interest (COIs). The process of redistricting aims to draw electoral boundaries in such a way as to keep COIs intact within electoral districts, rather than splitting them. Though the debate on how to define such communities is ongoing in the scientific literature on redistricting~\citep{webster2013reflections,forest2004information,malone1997recognizing}, states already employ a variety of proxy measurements~\citep{redistrictingCriteria} to describe how well underlying communities are captured by districts. Here, we review the problems associated with some of these common proxy measurements.

Compactness metrics, generally speaking, assume that communities form and spread in a geometrically-compact way over space \citep{stephanopoulos2012redistricting, duchin2018discrete}. While this is reasonable to assume in a general sense, there is no reason to assume that the exact degree of compactness calculated by a given metric should correlate to an exact degree of community coherence. Said another way, compactness can only be said to be a proxy for underlying communities, since it is not actually measuring community directly. Another commonly-used measurement to evaluate redistricting plans is the counting of municipal splits, which relies on the assumption that existing municipal boundary shapes capture underlying communities in a significant way \citep{winburn2008realities}. 
However, as with compactness measurements, this method fails to directly measure a particular aspect of the underlying communities, such as shared economic activity or religious affiliation \citep{gimpel2020conflicting}.

In response to the issues with the proxy metrics described above, researchers in political science have proposed several metrics that try to directly measure community coherence in a quantitative way. \cite{makse2012defining} argued that voters themselves should define COIs through ballot initiatives. \cite{stephanopoulos2012redistricting} introduced the concept of a territorial community test, which evaluates districts based on their spatial and socioeconomic congruence, as well as subjective affiliation to a given community. In a follow-up work, \cite{stephanopoulos2012spatial} described a metric called spatial diversity, which they use to measure spatial and socioeconomic congruence. To calculate spatial diversity, factor analysis is performed on census data to derive composite factors for individual census tracts. Then, the weighted standard deviation of each districts' composite socioeconomic categories is calculated, using the percent of variance described by each of the composite factors as the weight.

% Algorithm
\subsection{Spatial Interactions in Community Detection}
In a separate but related branch of research, geographers have been incorporating new, finely-grained spatial interaction datasets into more traditional methods for finding and quantifying communities. 
Spatial interactions, broadly speaking, are movements over space that result from a human process, and can range from human mobility flows, captured by surveys~\citep{nelson2016economic}, mobile phone calls or location device tracking data \citep{ratti2010redrawing,kang2020multiscale,xu2022understanding}, to commodity and information flows \citep{haynes2020gravity,li2020estimation,rao2022measuring}. Spatial interaction communities, then, are communities defined by a particularly dense set of spatial interactions.
As an example of this, \cite{gao2013discovering} used a mobile phone call dataset to measure human mobility flows across a city in China, and then used the resulting flows to identify spatial interaction communities. \cite{DONG2015278}, used mobile phone movement data for traffic zone division.
Building on these works, \cite{liang2022region2vec} used a spatial interaction flow ratio to evaluate communities found through various community detection algorithms, where the flow ratio is the sum of intra-community flows divided by the sum of inter-community flows.

Here, we argue that such a ratio is a theoretically-sound way to quantify community coherence for the purposes of redistricting, since spatial interactions are the embodiment of people's interests, whether social, economic, political, or otherwise. When people move in the same spaces, they are demonstrating involvement in the same geographic community.

In particular, we present this metric under the name \textit{interaction ratio} (IR), a new metric for redistricting plan evaluation which compares spatial interactions within districts to spatial interactions between districts in a global way. Essentially, this is a global metric to describe community strength, i.e., the degree to which district boundaries are aligned with the underlying spatial interaction communities.

\subsection{Redistricting Algorithms}
In the contexts described above, spatial interactions are used as inputs for community detection algorithms, which produce district assignments for a set of geographic units by grouping regions with dense spatial interactions together. In the redistricting literature, a broader set of algorithms called \textit{regionalization} algorithms are used for similar purposes, namely: 1) for the production of optimal plans, and 2) for the evaluation of proposed plans. In the former category, different techniques are used to produce a single districting plan that optimizes some set of criteria (e.g., compactness and fairness), while meeting constraints (e.g., equal population and geographic contiguity). Some examples of this approach include REDCAP \citep{guo2008regionalization}, the Divide and Conquer algorithm \citep{levin2019automated}, the Simulated Annealing Genetic algorithm \citep{bergey2003simulated}, and iRedistrict, a geovisual analytical program which incorporates graphical user input into a Tabu-style algorithm \citep{guo2011iredistrict}. The main problem with this class of optimization algorithms is that the redistricting problem is NP-hard, making the task of finding the globally-optimum plan computationally intractable for most political redistricting situations \citep{altman1997computational}.

Rather than searching for the global optimum, the other category of redistricting algorithms is focused on producing \textit{ensembles} of diverse plans that can be used to evaluate a given plan, where the ensembles are made up of representative samples from the space of valid plans. \textit{Outlier analysis}, in particular, compares a given plan to the distribution of values from an ensemble in several dimensions, such as fairness and compactness, and evaluates if the plan is an outlier or if it is relatively neutral. This type of analysis has been used successfully in several legal challenges to state-level redistricting plans, and is likely to see continued use in redistricting litigation \citep{ramachandran2018using}. Importantly, the generation of plans in this way allows for outlier analysis, but is not used to produce actual redistricting maps. This is for two reasons. 1) District maps produced in this way, while valid according to the constraints imposed on the algorithm, might not reflect practical considerations or all legal criteria, such as the requirement to maintain district cores, and 2) gerrymandering is a political process, and so legislators are incentivized to produce maps that favor themselves or their party, which is generally approached as an optimization task \citep{stephanopoulos2015partisan,brunell2008redistricting}.

In order to produce sample distributions, researchers from many domains have turned to Markov Chain Monte Carlo (MCMC) methods \citep{diaconis2009markov}. In a redistricting context, MCMC methods sample from some target distribution on the space of partition plans, perhaps defined by certain population and compactness thresholds, by making iterative changes to some initial district plan. The sample distribution approaches the target distribution as the number of iterations grows, with each iteration producing one map for the ensemble. Uses of this methodology for outlier analysis in redistricting be found in \cite{herschlag2017evaluating}, \cite{chikina2020separating}, and \cite{fifield2020automated}. Calculating the number of steps necessary to get a representative sample from the target distribution is impossible for most redistricting tasks, and so \cite{deford2019recombination} suggest using convergence heuristics and sensitivity analysis to demonstrate the robustness of the methods and settings used in a given Markov chain.

\subsection{ReCom Algorithm for Plan Comparison}\label{sec:plancomp}
One recent MCMC method that has proven useful for redistricting is the Recombination (ReCom) Markov chains algorithm \citep{deford2019recombination}.
Rather than using a typical flip-walk, which ``flips'' the assignment of a border unit from one district to another, the ReCom algorithm first merges two adjacent districts and then repartitions them by cutting a ``spanning tree'' (an acyclic subgraph in which all of the nodes are connected with the minimum possible number of edges) formed over the nodes (e.g., electoral precincts or wards) of the two merged districts. This method, compared to traditional methods, greatly reduces the time needed for convergence to a steady state (i.e., the mixing time), since the boundaries between adjacent districts are completely destroyed when they are merged, allowing for much more efficient movement throughout the space of valid maps. Examples of ReCom chains with state-level districts in the literature show that chains run for as little as 10,000 steps are able to produce well-mixed ensemble distributions, while traditional methods frequently do not pass these tests even after millions of steps \citep{becker2021computational,deford2019recombination}.

With an ensemble of plans in hand, a given map can be evaluated against the ensemble distribution for any number of redistricting metrics, allowing for multi-criteria assessment. Here, the term \textit{metric} simply refers to the particular measurement used to quantify a given criterion (e.g. Polsby-Popper compactness is a metric to quantify the criterion of district compactness; IR is a metric to quantify the communities of interest; efficiency gap is a metric to measure the fairness of a redistricting plan). The combination of ensembles of alternatives and multi-dimensional evaluation aids the process of redistricting in two ways:
1) by producing valid maps on actual geometry, we avoid the well-documented issues with the single-number ideals proposed by different metrics~\citep{deford2021implementing,bernstein2017formula,katz2020theoretical}. Rather than trying to come up with universal rules that hold any and everywhere, or under only a very specific set of conditions, comparing maps against ensembles of sampled maps allows for comparisons against actual alternatives for the area in question;
2) by using multiple criteria to evaluate district plans, we reduce the risk of over-relying on a particular criterion, which can easily misrepresent the ends or means they propose to support or measure, and embrace the complex, qualitative task of using various quantitative measures to support the pursuit of human ends, such as fairness, as redistricting objectives~\citep{duchin2021political}. 

Finally, although we use the ReCom algorithm as our base sampling algorithm in the research presented here, we also introduce new modifications (biased spanning tree and min-/max-cut, see section 3.4) that bias the sample distribution of spatial interaction ratio values towards higher or lower values to explore the redistricting plans with spatial interaction flow-connected communities.

\subsection{Educational Tools for Redistricting}
As can be seen in the previous sections, the process of redistricting plan production and evaluation permits an enormous number of options in every step, from setting compactness bounds to picking a evaluation metric. To help educate and engage the public about this complicated process, a number of open web-maps have been developed and made available online. One prominent example is Dave's Redistricting website, commonly referred to as DRA (\url{https://davesredistricting.org}). The website enables exploration of Congressional, State Senate, and State House districts for each U.S. state. Demographic and voting data are provided for different census boundaries, and there are plots of statistical analyses for common redistricting metrics. The web-map portion allows for basic map interactions, including pan and zoom, as well as the creation of new redistricting plans. Districtr (\url{https://districtr.org}) and DistrictBuilder (\url{https://www.districtbuilder.org}) are similar tools which offer similar capabilities. One feature lacking, however, in all of these tools is the ability to compare redistricting plans side-by-side, which would facilitate learning by the end-user by allowing them to visually find differences between maps~\citep{guo2011iredistrict}. For the purposes of this study, maps with different properties, like a high interaction ratio or compactness value, could be compared in such a way, allowing the user to understand how metric values impact map design. Additionally, if supported with statistical summaries, a web map of redistricting plans would allow for the interaction ratio to be understood in the context of existing metrics. To support these learning goals, we introduce an interactive web-map that allows for multi-criteria evaluation of different redistricting plans.

\section{Methods}
\subsection{Spatial Interaction Dataset Preparation}
In order to calculate the spatial interaction ratio, we employ SafeGraph Neighborhood Patterns (\url{https://docs.safegraph.com/docs/neighborhood-patterns}) as our underlying dataset of spatial interactions. 
This dataset records the number of anonymized cell phone ``visits'' from one location to another, where the place visits are recorded at the census block group (CBG) level for each temporal period (e.g., hourly, daily, weekly, and monthly) of the year. To protect privacy, SafeGraph employs differential privacy techniques on the flow dataset, including the removal of flows where counts are very low (i.e., less than 4 devices). In terms of the representativeness of SafeGraph data, the mobile phone users detected by SafeGraph represent about 10\% of the mobile phone users across the United States, with local variation, and therefore some scaling is required to estimate the population-level human mobility flows. Following \cite{kang2020multiscale}, we infer the population-level human mobility flows ($pop\_flows$) between a given origin and destination (O-D) CBG pair using the ratio of origin CBG population to the number of SafeGraph devices in the origin. The equation is as follows:\\
\begin{equation}
pop\_flows(o, d) = device\_flows(o, d) \times \frac{pop(o)}{num\_devices(o)}
\end{equation}

% location of flows used in calculation Redistricting/flow_mapper_2020_dev.ipynb
Using this equation, we take the Neighborhood Pattern dataset for every month of 2020 to produce an O-D flows matrix of the average monthly population flows between and within Wisconsin CBGs, where the $i_{th}$ row and $j_{th}$ column intersection is the average monthly population-level flow counts going from the origin CBG to the destination CBG. Although there are some human mobility pattern changes associated with the COVID-19 lockdown period \citep{gao2020mapping}, the monthly average OD flows over the year are relatively stable.

\subsection{Interaction Ratio and Comparison Metrics}\label{sec:metrics}

The interaction ratio (IR) is a global measure of spatial interaction community strength for a given redistricting map, where a spatial interaction community is defined by people moving within the same set of CBGs, and community strength is the degree to which district boundaries place geographic units (e.g., census tracts or CBGs) which have a relatively high number of spatial interactions between them within the same district.
Specifically, the degree to which a set of district boundaries encapsulates the underlying communities is calculated as the sum of district intra-flows divided by the sum of district inter-flows.
Intra-flows are defined as population flows $s_{o,d}$ where the origin CBG $o$ and the destination CBG $d$ are assigned to the same district (i.e., $D_{o} = D_{d}$).
Inter-flows are defined as population-flows originating from CBGs in one district and ending in another (i.e., $D_{o} \neq D_{d}$).
Therefore, a given set of districts generated by the ReCom algorithm (section \ref{sec:Recom}) has a higher IR if more flows originate and end in the same district relative to the number of flows that originate in one district but end in another. The following equation expresses this idea concisely, where $K$ is the total number of districts:

% \begin{equation}\label{eq:interactionRatio}
% \textit{IR} = \frac{\textit{Sum of Intra Flows}}{\textit{Sum of Inter Flows}} 
% % =
% % \frac{\displaystyle\sum_{i=1}^{k}\displaystyle\sum_{o=1}^{n_{dis=i}}{\displaystyle\sum_{d=1}^{n_{dis=i}}flows_{o,d}}}
% % {\displaystyle\sum_{i=1}^{k}\displaystyle\sum_{o=1}^{n_{dis=i}}{\displaystyle\sum_{d=1}^{n_{dis\ne i}}}flows_{o,d}}
% =\frac{\displaystyle\sum_{i=1}^{g}\displaystyle\sum_{o=1}^{n\in i}{\displaystyle\sum_{d=1}^{n \in i}flows_{o,d}}}
% {\displaystyle\sum_{i=1}^{g}\displaystyle\sum_{o=1}^{n\in i}{\displaystyle\sum_{d=1}^{n\in K}flows_{o,d}}}
% \end{equation}
\begin{equation}\label{eq:interactionRatio}
\textit{IR} = \frac{\textit{Sum of Intra Flows}}{\textit{Sum of Inter Flows}} 
=
\frac{\sum_{D_{o}=D_{d}}s_{o,d}}{\sum_{D_{o}\neq D_{d}}s_{o,d}}; D_{o}, D_{d} \in 1,2, \dots, K
\end{equation}

\begin{figure}[H]
    \centering
    \includegraphics[width=0.5\textwidth]{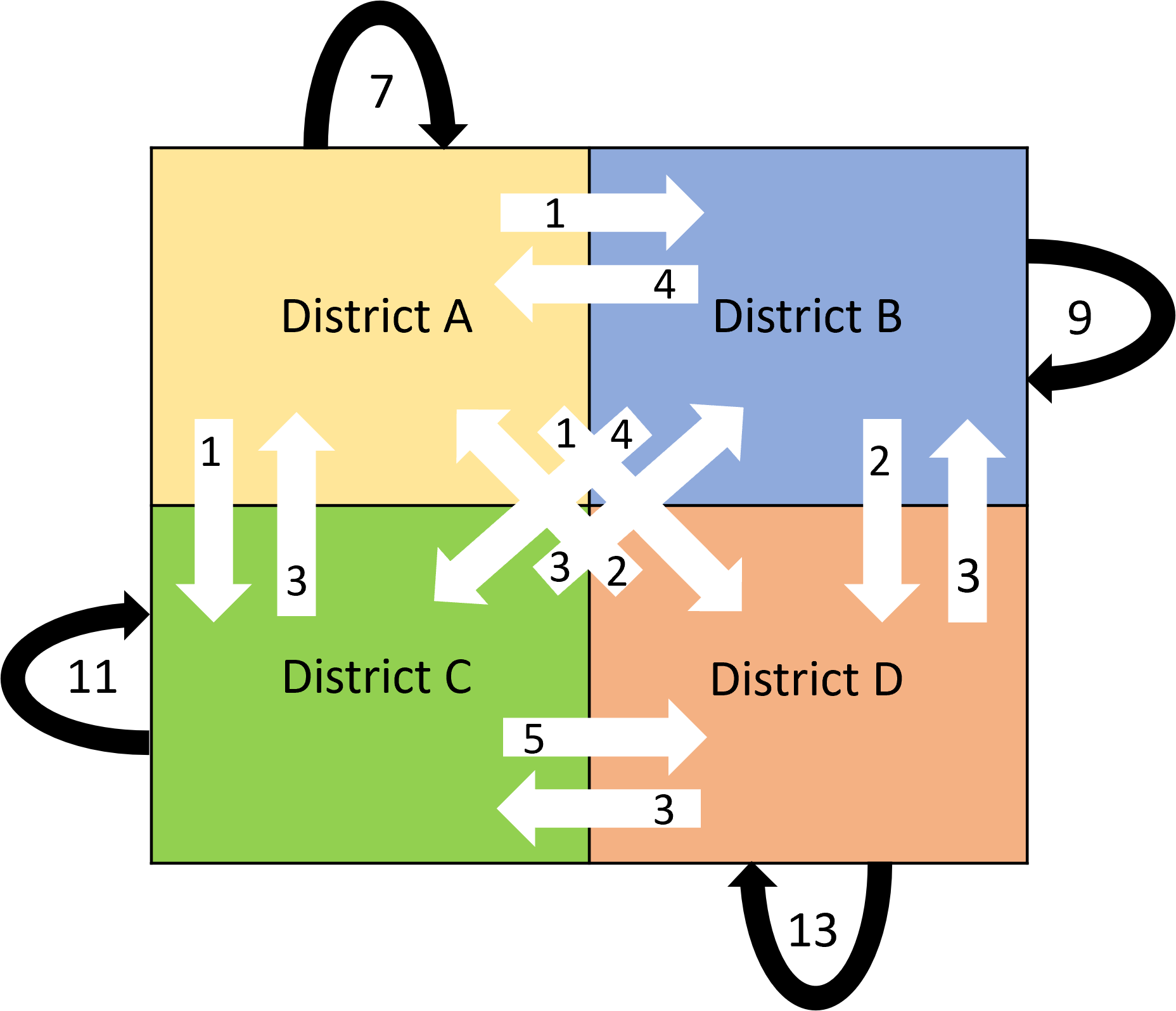}
    \caption{Intra and inter-flows for an example set of districts. Black arrows depict intra-flows, while white arrows depict inter-flows. [one-column width]}
    \label{fig:exampleFlows}
\end{figure}

To help demonstrate the calculation of the IR, we provide Figure \ref{fig:exampleFlows} using a hypothetical region, which shows the intra and inter-flows for a set of four districts. In this example, the underlying CBGs that comprise each district are not depicted so as to not overly-clutter the diagram. Instead, we show only the intra-flows for CBGs assigned to a given district, depicted with black arrows, and the inter-flows for CBGs assigned to different districts, depicted with white arrows. As seen in Equation \ref{eq:demoInteractionRatio}, the sum of intra-flows is the sum of black arrow flows, and the sum of inter-flows is the sum of white arrows flows, yielding an IR of 1.25 for this particular set of districts. Note that the IR is only one of the simple metrics to quantify the spatial interaction strength within communities. A normalized version of IR (i.e., the ratio of intra flows to total flows) can also be used to get a value between 0 and 1.

\begin{equation}\label{eq:demoInteractionRatio}
\textit{IR} = \frac{7+9+11+13}{1+1+1+4+4+2+3+3+5+2+3+3} = \frac{40}{32} = 1.25
\end{equation}
% \begin{equation}\label{eq:demoInteractionRatio}
% \textit{Interaction Ratio} = \frac{\textit{Sum of Intra-Flows}}{\textit{Sum of Inter-Flows}} = \frac{7+9+11+13}{1+1+1+4+4+2+3+3+5+2+3+3} = \frac{40}{32} = 1.25
% \end{equation}

% \\ describe heuristics for convergence and evidence about seeds and number of iterations
To measure how the IR varies with common redistricting metrics, we also calculate compactness and efficiency gap scores for the districts produced by the ReCom algorithm (section \ref{sec:Recom}). For compactness, we use the Polsby-Popper measure (Polsby and Popper, 1991) which provides a score in the interval of $[0,1]$, with 0 being the least compact and 1 being maximally compact. The equation is as follows:
% Polsby-Popper\ Compactness=4piArea/(Perimeter upcarrot 2 ) 
\begin{equation}\label{eq:pp}
\textit{Polsby-Popper Compactness} = 
\frac{4\pi Area}{Perimeter^{2}}
\end{equation}

In contrast to this area-perimeter ratio definition of compactness, the ReCom algorithm uses a discrete definition of compactness, where the compactness of a given plan is bounded by $n$ times the number of cut edges in the original plan, and $n$ is a multiplier set by the user. However, we report specific values of compactness using the Polsby-Popper metric, as it is a more common measurement of compactness in redistricting evaluation.

The efficiency gap is a measure of election fairness where wasted votes for the two major parties are counted for each district and then summed across districts \citep{stephanopoulos2015partisan}. Wasted votes for the winning party in a given district are votes cast in excess of what is needed to win the election (i.e., every vote after 50\% of votes plus one), while all votes for the losing party in a given district are considered wasted. Therefore, in this work the efficiency gap is calculated using votes from all districts with the following formula:
% Efficiency Gap=(Democratic Wasted Votes-Republican Wasted Votes)/(Total Votes)   
\begin{equation}\label{eq:eg}
\textit{Efficiency Gap} = 
\frac{\textit{Wasted Votes for Democrats}-\textit{Wasted Votes for Republicans}}{\textit{Total Votes}}
\end{equation}
The ReCom algorithm implementation and calculations for Polsby-Popper compactness and the efficiency gap are performed with the GerryChain Python package (\citeyear{gerrychainPkg}), while all other calculations are done in Python using standard libraries.

\subsection{ReCom Algorithm for Plan Sampling} \label{sec:Recom}
\begin{figure}[H]
    \centering
    \includegraphics[width=1.0\textwidth]{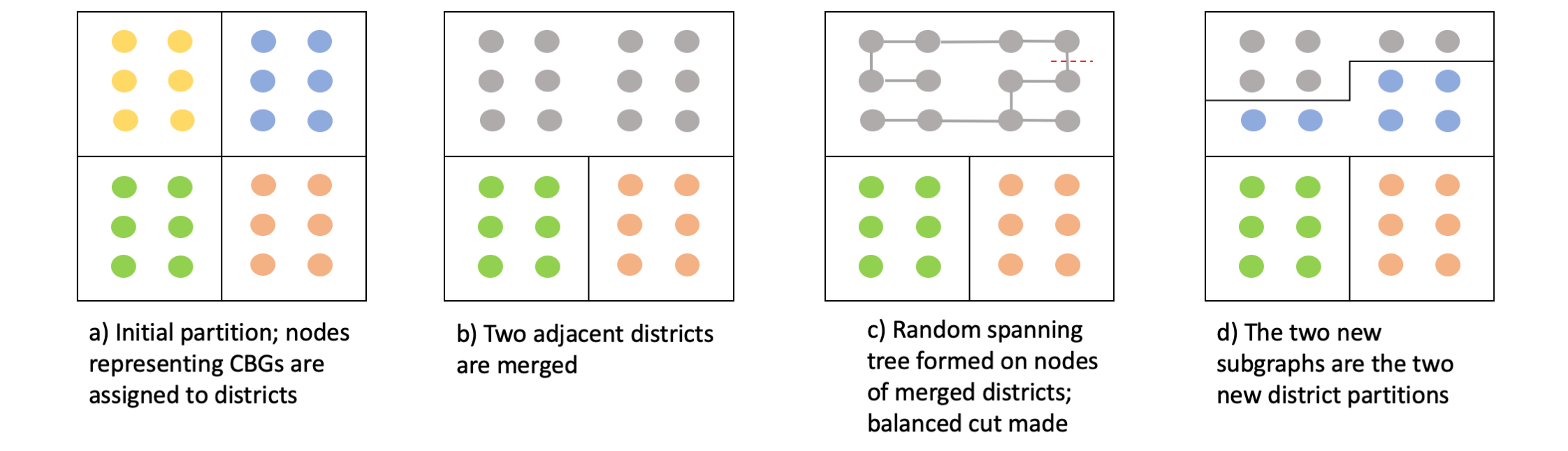}
    \caption{Demonstration of the ReCom algorithm on an example set of nodes (representing geographic units, e.g. census block groups). District assignments are represented by node color, and district boundaries are represented by black lines. The spanning tree formed on the nodes from the merged districts is represented in gray, and the cut to the spanning tree is represented with the dashed red line. [two-column width]}
    \label{fig:graphicalrecom}
\end{figure}

\begin{figure}[H]
    \centering
    \includegraphics[width=1.0\textwidth]{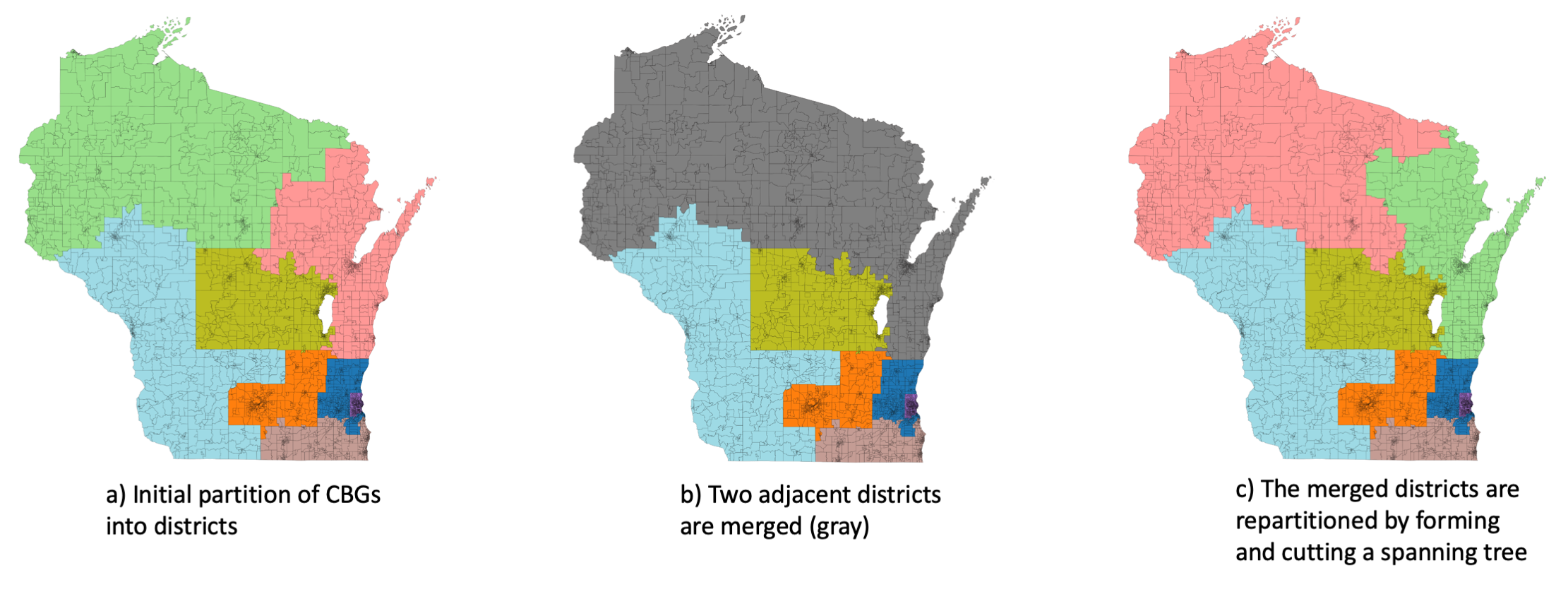}
    \caption{An example of Wisconsin  congressional districts being merged and repartitioned with the ReCom algorithm, using census block groups as the underlying spatial units. The pink and green districts are merged to form the gray district, and which in turn is repartitioned into two new districts. [two-column width]}
    \label{fig:wimerge}
\end{figure}

As shown in Figure \ref{fig:graphicalrecom}, the ReCom algorithm draws and cuts spanning trees on merged districts in order to efficiently repartition districts and sample possible arrangements. Since more compact districts allow for the creation of more spanning trees, compact districts are chosen with much greater frequency \citep{procaccia2022compact,deford2019recombination}. Rather than producing a uniform distribution of district arrangements, then, which would be dominated by non-compact districts, the ReCom algorithm targets the \textit{spanning tree distribution}, which refers to the subset of relatively compact district plans that get sampled when the spanning tree method is used in ReCom.
More specifically, the algorithm works in the following way (Figure \ref{fig:graphicalrecom}): a) there exists an initial partition in which nodes (e.g.,  electoral precincts, census block groups or other geographical units) are assigned to different districts; b) when two districts' nodes in the graph are merged, a random spanning tree is built on the induced subgraph's edges; c) cuts are then explored to see which would maintain (near) population equality between the two new partitions; d) a cut is made, creating the new districts. An example of this process with state-level geometry is shown in Figure \ref{fig:wimerge}.

\subsection{ReCom Algorithm Modifications for Biased Sampling}
We note that some weighting scheme for tree or cut-selection could potentially change the sample distribution of districting plans, making this an ideal place for intervention to build upon ReCom's useful properties. In particular, for trying to preferentially sample districts that better capture the underlying spatial communities, as described by the IR, we could choose cuts or trees that favor districts with higher IR values. Modifying the sample distribution of districting plan maps in these ways is important for the same reason that generating an ensemble of districting plan maps is important in the first place: having a distribution of maps allows one to evaluate a given map against a set of alternatives. Being able to modify the target distribution with respect to IR would give redistricting researchers some ability to specify the acceptable target distribution of redistricting plans.

As the focus of our experiments, we produce and analyze the 2020 congressional district maps for the state of Wisconsin.
Our goal is to compare trade-offs between compactness, the efficiency gap, and our proposed IR metric for quantifying community strength, and so sampling as widely as possible for each respective metric is desirable. To that end, we allow for up to a 3\% population deviation between the largest and smallest districts. 
For compactness exploration, we consider high and low compactness limits, defined as 1 or 5 times the number of cut edges in the initial plan. With these experimental settings in place, we next describe the ReCom algorithm and our modifications to it.

The steps of the algorithm are as follows (adapted from \cite{deford2019recombination}):
\begin{algorithm}[h]
\caption{Recombination (ReCom)}\label{alg:cap}
\textbf{Input:}\text{ Dual graph $G = (V, E)$, and the current partition $P=W_1,\ldots, W_k$ on $G$} \\
\textbf{Output:}\text{ The next partition $Q$} 
\begin{algorithmic}
\State Select two adjacent districts, $W_i$ and $W_j$, from $P$
% \textbf{Try:}\\
\State Form the induced subgraph $H$ of $G$ on the nodes of $W = W_i \cup W_j$
\State{Form a uniform random spanning tree $T$ on $H$}
\While {There is no edge in $T$ that creates contiguous districts with equal population}
\State{Replace $T$ with a new random spanning tree on $H$}

\EndWhile
\State{Pick a valid cut uniformly at random}
\State \Return{$Q$}
\end{algorithmic}
\end{algorithm}

To both see the range of IR values possible within the spanning tree distribution of  congressional district plans and to explore how to most effectively sample from such a range, we employ two methods.

In the first method, we adapt how the random spanning tree (RST) is built on the nodes of the merged districts. While the original method builds a maximum spanning tree on the induced subgraph, whose edge weights have been randomized, our method introduces two additional steps to this process: 1) after every edge is given a random weight from a uniform distribution on the interval $[0,1]$, we then multiply that edge by the actual population flow weight (i.e., the true edge weight), and 2) we multiply this value again by a bias term. Then, as in the original method, a maximum spanning tree is built with these weights and a random cut is chosen, subject to the population and compactness constraints. Essentially, this method seeks to preferentially build spanning trees that keep neighbors with high flow weights between them within the same district. For exploring higher and lower IR values, we can modify the sign and magnitude of the bias term.
We will refer to this method as the \textit{biased RST} method. In Figure \ref{fig:graphicalrecom}, this method would impact the third pane, biasing the spanning tree built on the grey nodes such that neighbors with high flows would be more likely to be directly connected.

Our second method tries to bring a fuller picture of the IR into the selection of the partitions made. With this method, we use the standard ReCom method to draw a random spanning tree on the set of nodes from two merged districts, and produce a list of valid cuts. Then, we take this list of cuts and evaluate which one most (or least) improves the IR value of the map, and take that cut. As with the biased RST method, one of our goals with this second method is the exploration of possible IR values, and so we are interested in exploring district plans that have higher IR values as well as district plans that have lower IR values. Accordingly, we refer to the two methods as either the \textit{max-IR cut} or the \textit{min-IR cut}, depending whether the max or min cut was chosen. In Figure \ref{fig:graphicalrecom}, these methods would impact where the cut (represented with the red dashed line) is made on the grey random spanning tree.

The ReCom method introduced in \cite{deford2019recombination} picks a spanning tree uniformly at random from among all of the spanning trees of the merged districts. However, the GerryChain \citeyearpar{gerrychainPkg} implementation of ReCom builds a random maximum spanning tree by default. Internal experiments revealed that the two methods yielded comparable results, but that building random maximum spanning trees was substantially faster, and so the default method is used here. For clarity, the standard ReCom runs without any bias sampling or modifications will be referred to as \textit{RST}, referring to the method that builds random maximum spanning tree.

% different experimental settings - table?
\subsection{Experimental Settings for Baseline and Modified ReCom Algorithms}\label{sec:settings}
We take the redistricting plan produced by the People's Map Commission for Wisconsin 2020 Federal House Districts (\url{https://govstatus.egov.com/peoplesmaps/work-records}) as our seed map. As noted in \cite{deford2018comparison}, the ReCom algorithm results do not vary much with the seed map or sample path through the state space. Internal experiments for our research here found similar results, and so the experiments that we present here all use the same seed map. For geometry and demographic data, we employ the CBG-level data provided by the Redistricting Data Hub (\url{https://redistrictingdatahub.org/dataset/wisconsin-block-pl-94171-2020/}). To get voting data that is comparable with the human mobility flows, we disaggregate ward-level election results to the CBG-level based on the number of citizens that are at least 18 years old in age, as is a standard practice in the Political Science literature (see \cite{benade2021ranked} and \cite{becker2021computational} for examples). Votes cast for Democrats and Republicans in the 2018 Federal Congressional election are used to calculate the efficiency gap.

Using these inputs, we run a total of six chains to 1) test the biased RST and max/min-IR cut methods for sampling higher and lower IR values, and 2) sample a wide range of values for all metrics for use in the multi-criteria assessment of redistricting plans. Prior work by our group had shown that the distribution of IR values is largely dependent on compactness settings (the same pattern seen in our experiment results in section \ref{sec:resultsampling}), and so we break up the experiments into three chains which evaluate higher IR values, and three chains which evaluate lower IR values, as seen in Table \ref{Tab:experiments}.

\begin{table}[H]
\centering
\caption{Experimental settings for the six ReCom chains used to produce redistricting plan ensembles.}
\begin{tabular}{cc}
\hline
\textbf{Compactness Bound Multiplier} & \textbf{ReCom Method}               \\ \hline
1\!$\times$                                    & RST                           \\
1\!$\times$                                    & Biased RST (100 bias factor)  \\
1\!$\times$                                    & Max-IR cut                    \\
5\!$\times$                                    & RST                           \\
5\!$\times$                                    & Biased RST (-100 bias factor) \\
5\!$\times$                                    & Min-IR cut                    \\ \hline
\end{tabular}
\label{Tab:experiments}
\end{table}

As mentioned in section \ref{sec:plancomp}, 10,000 iterations has been shown to be sufficient for convergence with a number of state-level redistricting chain experiments, where each iteration involves two districts being merged and repartitioned. Internal work by our group found that even as few as 3,000 iterations produced relatively stable distributions, and going as high as 12,000 iterations did not continue to change ensemble distributions (stability was evaluated visually by inspecting bar plots of the distributions, similar to what is done in \cite{deford2019recombination}).
Therefore, each experiment is run for 10,000 iterations.

\subsection{Interactive Web Map Development}\label{interactivewebmap}

In addition to describing and analyzing the sampled redistricting ensembles (sections 4.1 and 4.2), we also present some of the results in the interactive web map. In particular, we present the three redistricting plans that maximize the IR, the efficiency gap, and compactness, respectively, such that each of those maps represents the extreme of a given metric for the various district samples drawn.
Furthermore, we provide demographic attributes on the district level for each of these maps. To encourage exploration as a learning method for the user, we present two side-by-side interactive maps, as well as a linked, scented widget, which cues users into the attribute values underlying districts. We also present the sampled ensembles of district plans in the form of an interactive 3D cube. Further down the web page, we provide linked text descriptions for each of the calculated metrics.
The web map itself (see section \ref{sec:webmap}) is developed using a technology stack including JavaScript, HTML, CSS, Leaflet, D3, and Map Sync libraries.

To demonstrate that our web map provides novel functionality in the space of public redistricting resources, we compare our web map to the popular redistricting websites described above, namely the Dave's Redistricting (DRA), Districtr, and DistrictBuilder. The functionalities examined are as follows:

\begin{itemize}
    \item Compare multiple plan boundaries simultaneously;
    \item Compare a selected plan to the distribution of plans;
    \item Compare metrics for multiple maps;
    \item Plan creation;
    \item Provide optimal maps for comparison;
    \item Provide redistricting metrics and demographics for districts;
    \item Provide reference layers;
    \item Provide map re-expression.
\end{itemize}

To assess the usability of the developed web map, we survey 10 users with GIS or web map experience in the University of Wisconsin-Madison to complete a series of six tasks related to map and plot reading. Overall, the tasks in the assessment evaluate if the web map design facilities the use of the main web map features, e.g. are users able to change which map is displayed in each of the map panes. University students with GIS or web map experience, i.e. at least one course in either area, are invited as the participants, as the intended audience for the interactive web map is researchers and members of the public engaged in the redistricting process, and some knowledge of mapping tools is common among these groups. The participants include two undergraduate and eight graduate students, none of whom had experience with political redistricting.
For each question, we report the average response accuracy, where responses are either correct or incorrect. The tasks are as follows:

\begin{enumerate}[A.]
    \item Add the CBG overlay to the left map pane;
    \item Hover over the district that contains the label (``Milwaukee") in each of the map panes;
    \item Change the 2022 Enacted Map panel to show the Max Compactness Map;
    \item Hover over the line in the parallel coordinate plot corresponding to the district with the highest $inter\_flows$ value for the Interaction-Ratio map;
    \item Read the district-level efficiency gap (i.e., the $Dis\_EffGap$ value) for district containing the city of Madison in the 2022 Enacted Map;
    \item Hover over the point with the minimum compactness value in the interactive redistricting values cube.
\end{enumerate}

Task A tests the usability to add a map overlay. Task B tests the usability to compare differences in district boundaries between two maps. Task C tests the usability to change which maps are displayed. Task D tests the usability for a user to read the parallel coordinate plot and associate its values with district boundaries. Task E tests the usability to use map boundaries to read values from the associated plots. Task F tests the usability to move the interactive 3D cube in order to read redistricting metric values. To complete the assessment, each participant is read and shown the task description, using one screen, and then they complete the task on another screen. The member of the research team administering the assessment reads the task and responds to any questions while sitting next to the participant. Questions regarding the meaning of the task are answered, while questions regarding the execution of the tasks, i.e. ``so I should click right here?" are not answered so as to not provide assistance to the participant beyond the affordances of the web map. Each assessment is performed individually. Task completion accuracy is recorded as 1 for correct or 0 for incorrect, with answers being collected audibly or visually by the researcher administering the test, e.g. ``the district-level efficiency gap value is -0.33 for the district containing Madison" and seeing that the participant's mouse is hovering over the correct district.

\section{Results}\label{sec:results}
\subsection{Baseline and Modified Sampling Distributions}\label{sec:resultsampling}
\begin{figure}[H]
    \centering
    \includegraphics[width=1.0\linewidth]{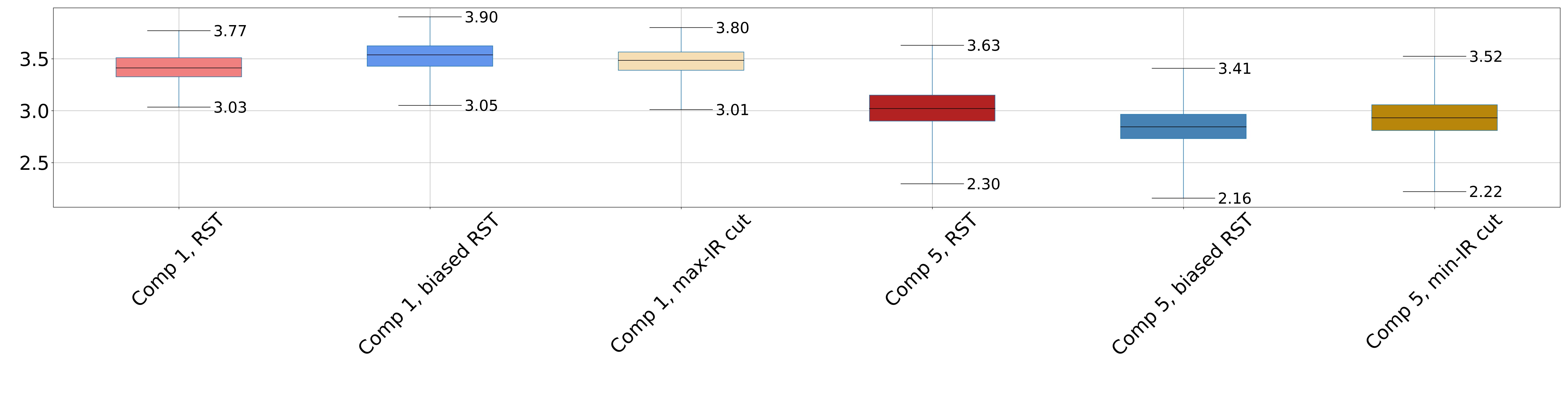}
    \caption{IR values (y-axis) for ensembles produced by RST (red), biased RST (blue), and max/min-cut RST (brown) algorithms. Each chain was run for 10,000 iterations. Lighter colors on the left are for the tighter compactness threshold of one, while darker colors on the right are for the looser compactness threshold of five. [two-column width] }
    \label{fig:boxplots}
\end{figure}

We first present the spanning tree sampling results from the ReCom algorithm and adaptions to it, with respect to IR values. As seen in Figure \ref{fig:boxplots}, the most important factor in the IR values sampled is the compactness, with the compactness limit of 1\!$\times$ the number of cut edges in the original map allowing for substantially higher IR values to be sampled when compared with a compactness limit of 5\!$\times$ the number of cut edges. Given that, we compare methods with respect to compactness thresholds. 

With a compactness limit of 1\!$\times$, the RST method sampled IR values ranging from 3.03 to 3.77. Both the 100 biased RST and max-IR cut methods produced distributions that were significantly different from baseline, with KS-tests comparing them to the RST method yielding $p$-values of 0.00. The biased RST method, in particular, was able to sample IR values as high as 3.90, while the max-IR cut method did not substantially increase the range over baseline.

Similar results were produced with a compactness limit of 5\!$\times$, with the -100 biased RST method producing a sample with both a lower range and center of IR values, compared to the standard ReCom method, and the min-IR cut producing similar but smaller improvements. The baseline RST method sampled IR values ranging from 2.30 to 3.63, while the biased RST method sampled values ranging from 2.16 to 3.41. KS-test values comparing the biased RST and min-IR cut methods to RST produced $p$-values of 0.00, suggesting that those samples come from different underlying distributions than that of the standard ReCom sampler.

While the biased RST and max-IR cut methods only shift the range of IR values a modest amount over the spanning-tree distribution, the ability to shift the distribution at all is important for several reasons. For one, legal challenges employing outlier analysis must specify the distribution that is being sampled from \citep{procaccia2022compact}, and the methods detailed here provide a way to sample districts that better match communities defined by spatial interactions. For another, the difference in performance between the two methods gives us some insight into what mobility flow factors play a larger role in changing IR values. The biased RST method, which was able to shift the distribution within the spanning tree distribution the most, works by randomly weighting all edges, multiplying those edges by the human mobility flows of each edge, respectively, and then building a maximum spanning tree on this new set of weights. This suggests that higher IR districts are, on average, sampled by trying to increase intra-flows between adjacent sub-geographies, rather than trying to decrease inter-flows or both increase intra-flows and decrease inter-flows at the same time, as in the max-IR cut algorithm. 

It may have been difficult to sample higher IR values because the ReCom algorithm is already producing samples near the upper range of possible values. Consider, for example, our seed map, the map produced by the People's Map Commission. The map was designed with the intention of keeping COIs intact (\url{https://evers.wi.gov/Documents/PMCCriteriaMemoFINAL.pdf}), and has an IR value of 3.04. Sampling lower IR values is likely possible, but undesirable, as producing very low-values IR plans would probably involve producing very non-compact districts which would not be used in redistricting.

For one final point, we also argue that even the modest increase in range for the IR values sampled between methods is important. For instance, consider the change in mobility flows from the highest IR value sampled by the ReCom algorithm 3.77, to highest value sampled by biased RST method, 3.90. Going from 3.77 to 3.90 (an increase of about 3\%) represents a change of 549,605 human travel trips being moved from inter-flows to intra-flows, which represents about 1\% of total flows in our dataset. While this is a modest amount, it is a tangible increase in community quality. Furthermore, the use of outlier analysis to challenge maps relies more on the center of the IR value distribution, rather than the range, and the sampling methods introduced here significantly shift the center of the sampled distributions.

\subsection{Multi-Criteria Assessment of Ensembles of Redistricting Plans}\label{sec:results3Dcube}

\begin{figure}[H]
    \centering
    \includegraphics[width=1.0\linewidth]{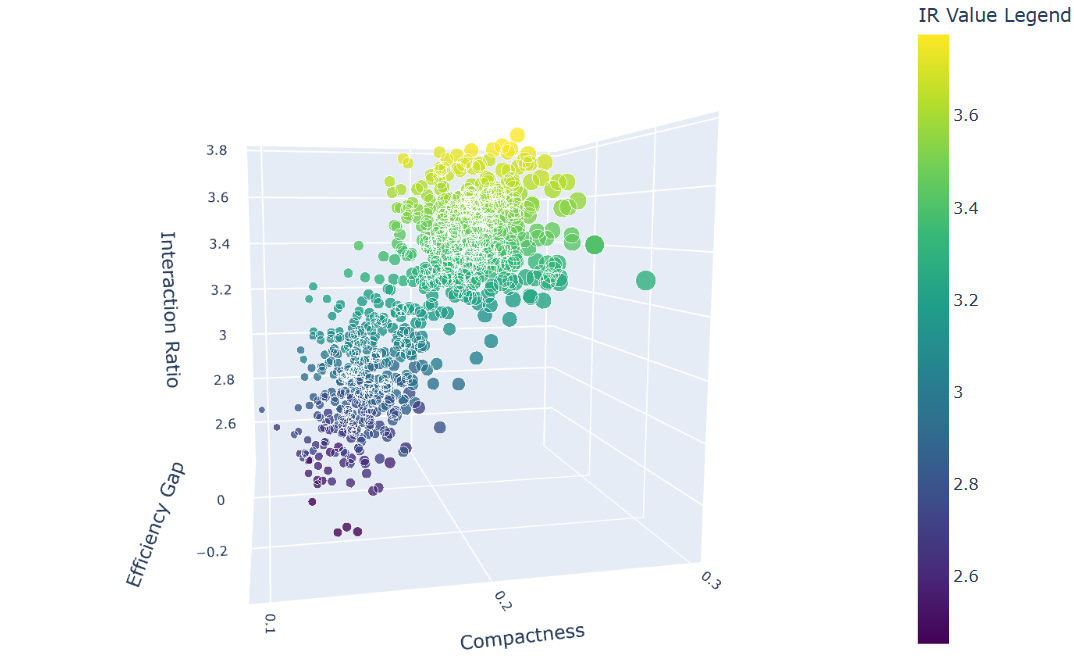}
    \caption{The 3D-cube visualization focused on the IR and compactness dimensions, where IR value is shown in color, and compactness is shown with marker size. Compactness has a monotonic relationship with the IR. [two-column width]}
    \label{fig:interaction_cube}
\end{figure}

Taking the ensembles of redistricting maps described in the previous section, we show the combined redistricting results in a multi-dimensional fashion, considering IR, compactness, and efficiency gap values. The goals of this analysis are to explore the space of valid redistricting plans and the trade-offs between the various criteria. Combining the sampled ensembles in this way allows us to explore a wider space of valid maps, but also means that the resulting distribution should not be used for outlier analysis. 
To highlight the relationships between the three metrics used, we produce a 3D-cube that plots sample values in three dimensions. While this cube is interactive in the web-map described below, we present two different views of the cube here to show the sampling results. 
In Figure \ref{fig:interaction_cube}, the 3D-cube is oriented to show the range of IR and compactness values found by the various ensembles of redistricting plans, with higher IR values colored in yellow, and lower IR values colored in purple. Higher compactness values are shown with larger dot markers. As seen in the figure, IR has a monotonic relationship with compactness, which is measured with the Polsby-Popper formula (eq. \ref{eq:pp}). Notably, a given compactness value can still allow for a modest range of IR values.

\begin{figure}[H]
    \centering
    \includegraphics[width=1.0\linewidth]{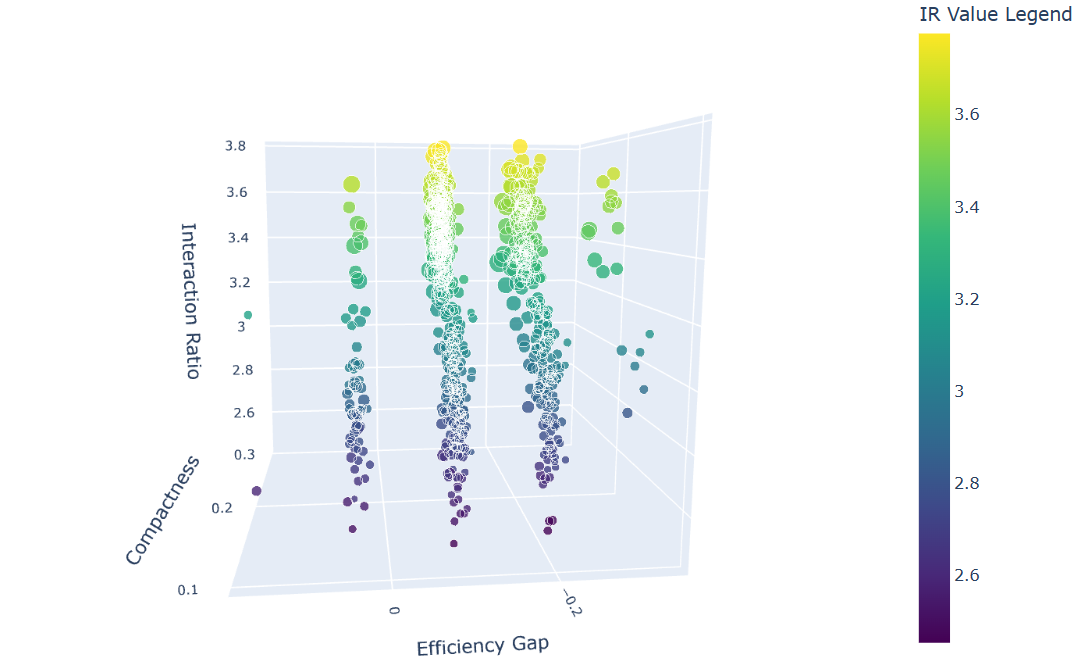}
    \caption{The 3D-cube visualization focused on the efficiency gap-dimension. The visible bands correspond to different seat allocations between the two major parties. Negative efficiency gap values indicate that more Democrat votes are wasted, and that Republicans gained a disproportionate number of seats relative to their votes. [two-column width]}
    \label{fig:eff_gap_cube}
\end{figure}

In Figure \ref{fig:eff_gap_cube}, we show the 3D-cube visualization focused on variability in the efficiency gap. Negative values indicate that more Democrat votes are wasted, and that Republicans gained a disproportionate number of seats relative to their votes, while positive values indicate the other way around. The majority of values are negative, showing that valid district plans generally favor Republicans in Wisconsin. 
All efficiency gap values, however, cluster into a few visible bands. This reflects that the efficiency gap is largely tied to seat allocation~\citep{duchin2018outlier,bernstein2017formula}, where each band represents another Congressional seat flipping from one party to another. The noise in each band is due to differences in voter turnout and population variation between districts, the latter of which is allowed only for sampling purposes (legally, districts must be of equal population). 

The fact that the majority of plans favor Republicans, even while Democrats received more statewide votes for the election used in this analysis, is generally considered to be a reflection of the partisan gerrymandering in Wisconsin political geography~\citep{chen2013unintentional,herschlag2017evaluating,wang2016three}: Democratic voters tend to naturally pack themselves in cities, namely Madison and Milwaukee, while Republican votes tend be more spread out in rural areas. In order to overcome this settling bias and produce maps that are more favorable to Democrats, districts encompassing rural areas nearby cities would essentially need to trade some of their rural area for a portion of city area, which often necessitates the creation of non-compact districts. What exact sorts of spatial clustering in political geography result in different partisan leans, however, is still an open question~\citep{duchin2021political,guo2008regionalization}. 

Indeed, appeals to partisan proportionality need to contend with how communities of interest are impacted by such practices. Using the ensembles sampled here, we can see the trade-offs between compactness, IR, and partisan bias by comparing low and high compactness biased RST chains, which together have the lowest and highest range of IR values, respectively, of any of the chains. Since compactness generally has a monotonic relationship with IR value, as seen in the 3D-cube visualization, we can consider these two chains to represent a low compactness and low IR ensemble, and high compactness and high IR ensemble, respectively. To compare these two ensembles of redistricting plans in terms of efficiency gap and seat allocation, we count the number of times in each ensemble, respectively, that a given seat allocation occurred. Then, we calculate the average efficiency gap for all of the district plans in each seat allocation category, with the resulting averages corresponding to the efficiency gap bands noted in the  3D-cube. For reference, statewide vote percentages for the election in use (US House 2018) were  $\sim$54\% and $\sim$46\%, for Democrats and Republicans, respectively, resulting in a slightly negative efficiency gap for district plans where equal seat allocation was achieved.

\begin{figure}[H]
    \centering
    \includegraphics[width=1.0\linewidth]{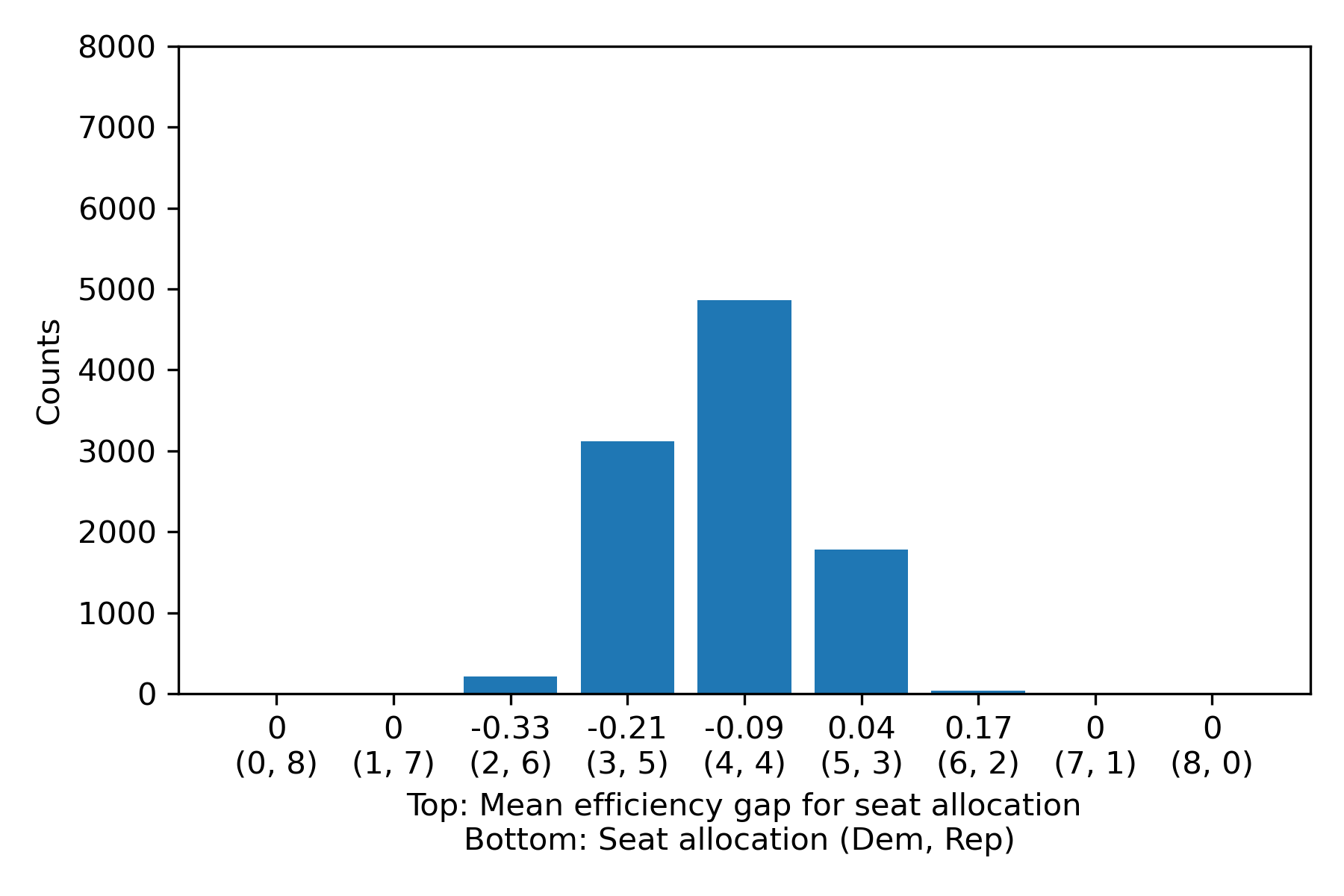}
    \caption{Negatively biased RST ensemble results, with low compactness and IR values, grouped by seat allocations. Seat allocation is reported on the lower level of labels on the x-axis in the format of ``(Democrat, Republican)" seats won under a given plan, e.g. the label ``(2,6)" means that Democrats won two seats and Republicans won six. The average efficiency gap is reported for each seat allocation ratio. [two-column width]}
    \label{fig:lower_ir_bar}
\end{figure}

\begin{figure}[H]
    \centering
    \includegraphics[width=1.0\linewidth]{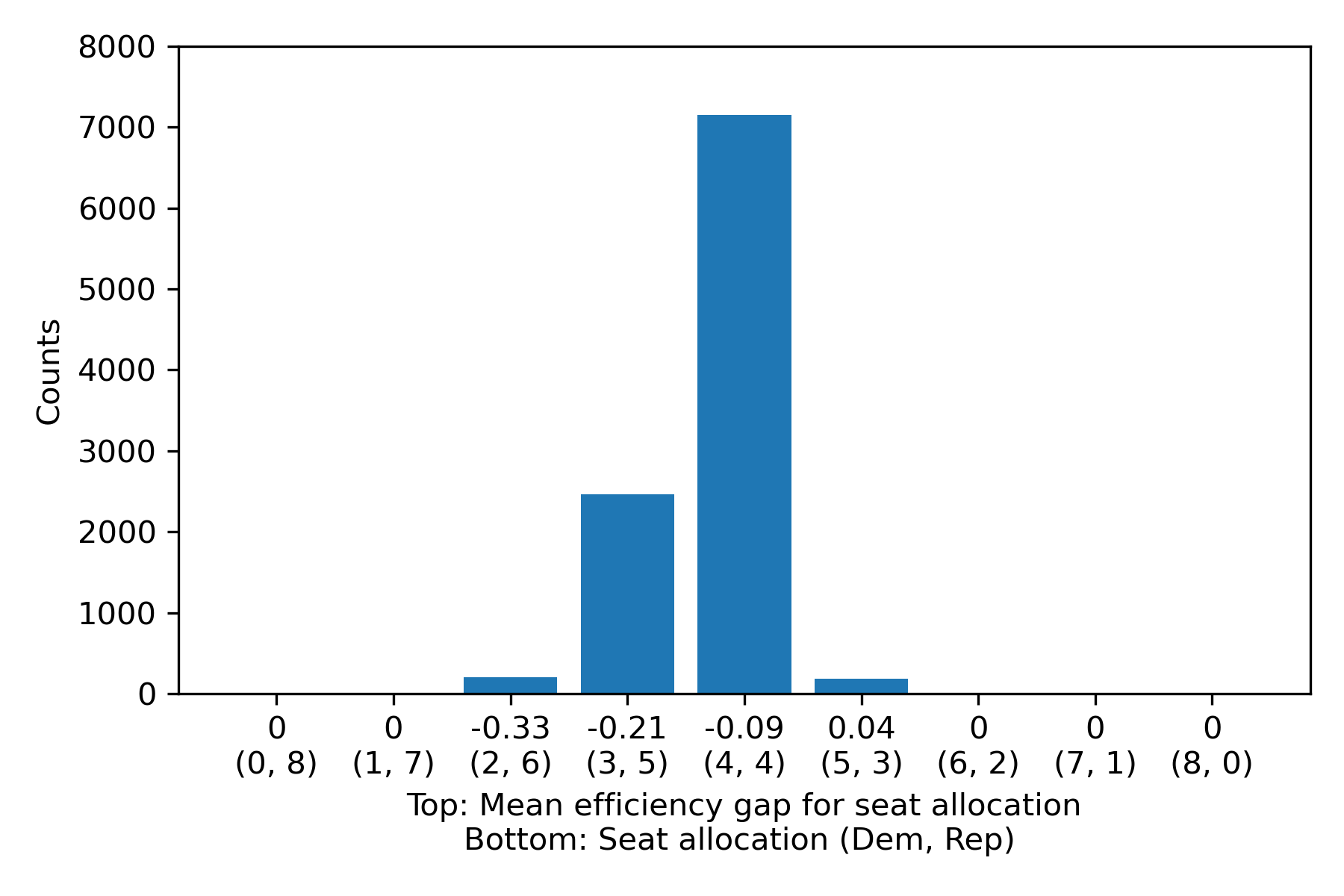}
    \caption{Positively biased RST ensemble results (with high compactness and IR values) grouped by seat allocations, where seat allocation is reported as Democrat, Republican seats won under a given plan. The average efficiency gap is reported for each seat allocation ratio. Compared to the negatively biased RST ensemble, this ensemble produced many more maps with proportional representation (4:4). [two-column width] }
    \label{fig:higher_ir_bar}
\end{figure}

As seen in Figure \ref{fig:lower_ir_bar}, the low compactness and low IR ensemble produced seat allocations varying between 2-Democrats, 6-Republicans (in total 8 seats) and the inverse, with most values falling in 3-5, 4-4, and 5-3 seat allocation categories. The \textit{proportionality} (4-4 seat allocation) is achieved in a relatively small majority of redistricting maps (compared to Figure \ref{fig:higher_ir_bar}), and the same category has a relatively neutral efficiency gap average.  In contrast, the high compactness and high IR ensemble produced a much larger fraction of maps that had proportional outcomes overall, as seen in Figure \ref{fig:higher_ir_bar}, suggesting that pursuing higher IR maps to better encapsulate COIs may also help produce results that have, on average, more proportional outcomes. Importantly, this analysis applies to the spatial scale used (US congressional districts composed of CBG-level geometry) and the particular region used (Wisconsin); the relationships between IR, compactness, and proportionality may differ for analyses at different spatial scales or in other regions.

\subsection{Interactive Web Map}\label{sec:webmap}

\begin{figure}[H]
    \centering
    \includegraphics[width=\linewidth]{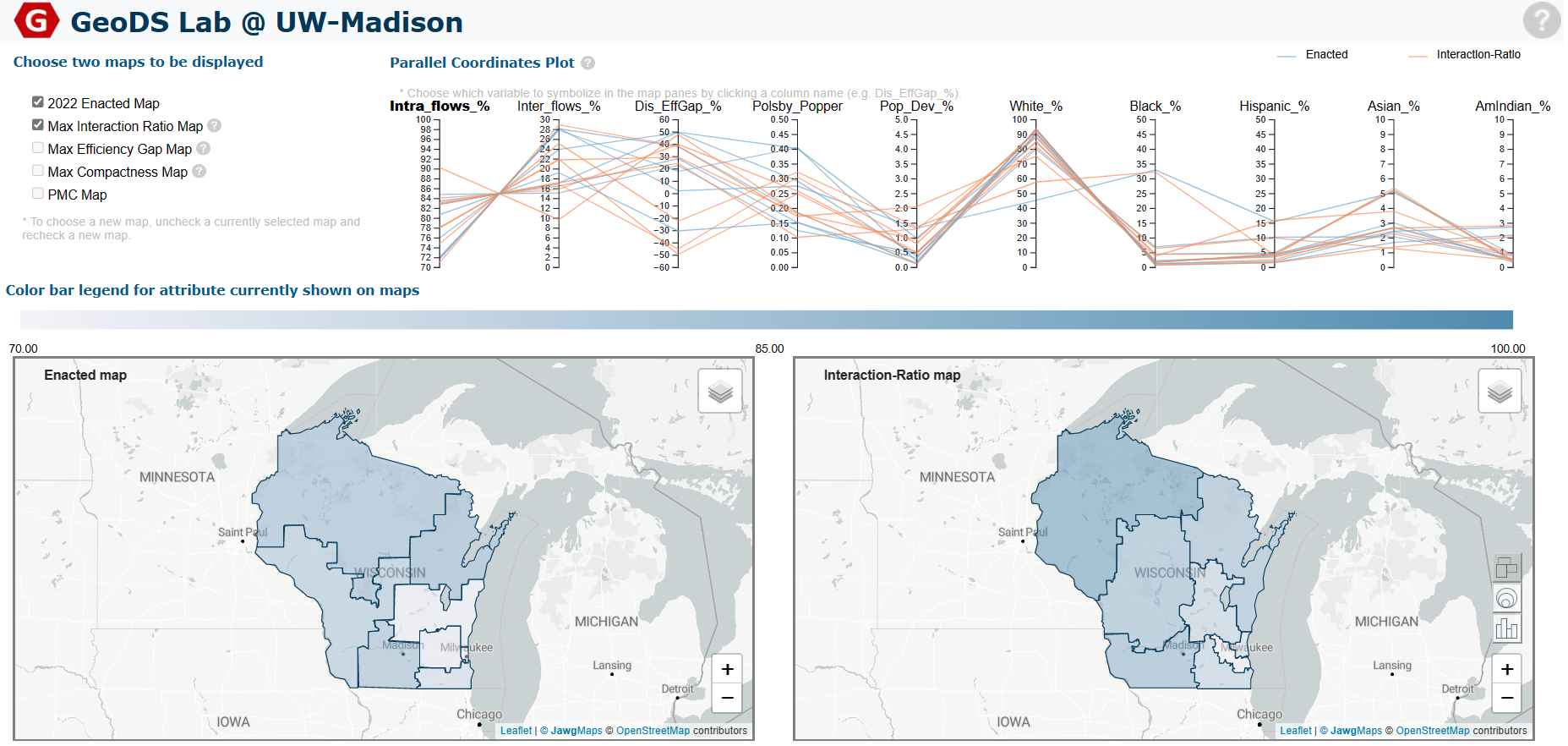}
    \caption{Our developed interactive web map, with the 2022 enacted district plan displayed in the left map panel, and the maximum Interaction Ratio map displayed on the right. [two-column width]}
    \label{fig:map_panes_r_votes}
\end{figure}

\begin{figure}[H]
    \centering
    \includegraphics[width=1.0\linewidth]{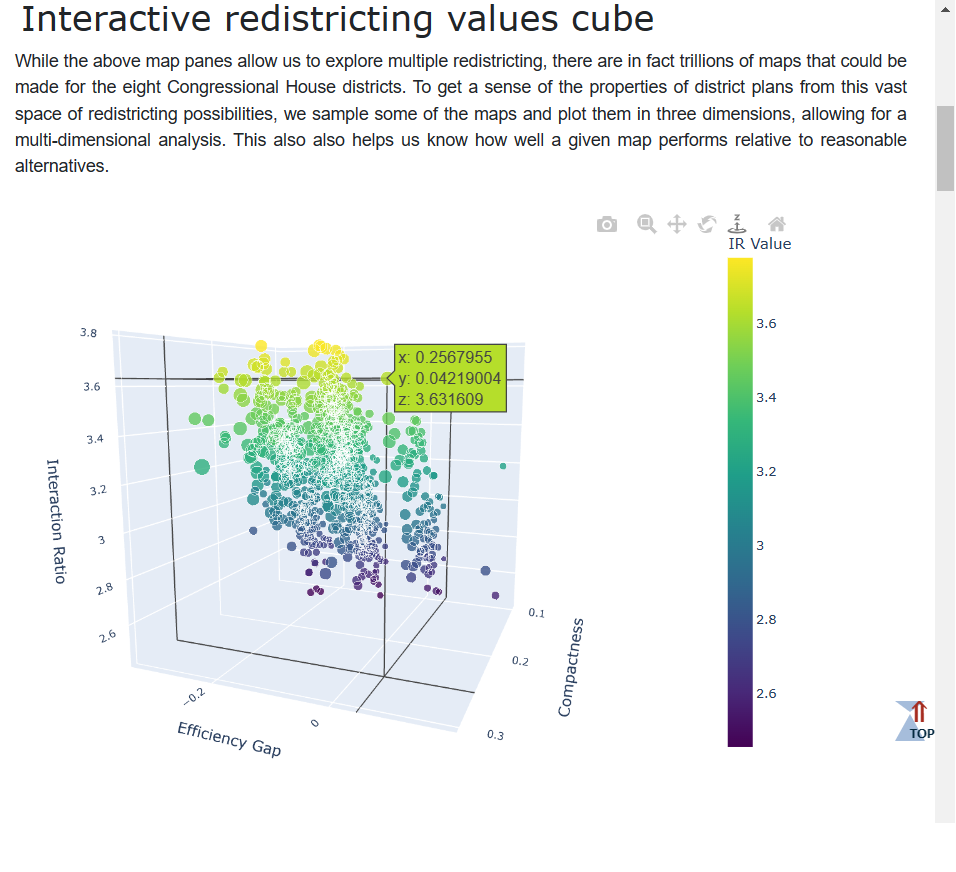}
    \caption{Variables x, y, and z on the label produced by mouse-over correspond to compactness, efficiency gap, and IR values, respectively. [two-column width]}
    \label{fig:cube_mouseover}
\end{figure}

In our developed web map, as seen in Figure \ref{fig:map_panes_r_votes}, public users are allowed to choose from Wisconsin congressional district plans that are optimized for compactness, the IR, and the efficiency gap, respectively. We also provide the currently enacted district map, and a districting plan map from the People’s Map Commission, for reference.  Existing works have shown the importance of demographic and socioeconomic variables for public policy making~\citep{kind2018making,hou2021intracounty}, and so demographic attributes and mobility flow values for each district can be visualized on the choropleth maps by selecting the corresponding header (with bold text) on the parallel coordinate plot. The same values can also be resymbolized as bar plots or proportional symbol maps. Finally, an OpenStreetMap base map provides spatial context by showing city and road names.
The two map panels are synchronized, such that the panning or zooming on one map is performed in equal measure on the other map. Map interaction, synchronization and the ability to select which variable to display are key features in making this web map educational. By exploring the redistricting maps and trying out different interactions on the website, users can get a better grasp of how district map attributes relate to each other.
Similarly, users can interactively reposition the 3D cube for the associated results of ensemble plans. An additional feature of the interactive form of the 3D cube is the ability to view point attributes, displaying the compactness, efficiency gap, and IR values for any point that is hovered over (Figure \ref{fig:cube_mouseover}). We also include text descriptions and diagrams below the interactive maps and charts, providing redistricting context and implementation details for the evaluation metrics used in the web map. The map is available to the public at \url{https://geography.wisc.edu/geods-file/redistricting/}.

\subsection{Interactive Web Map Usability Assessment}
\begin{figure}[H]
    \centering
    \includegraphics[width=1.0\linewidth]{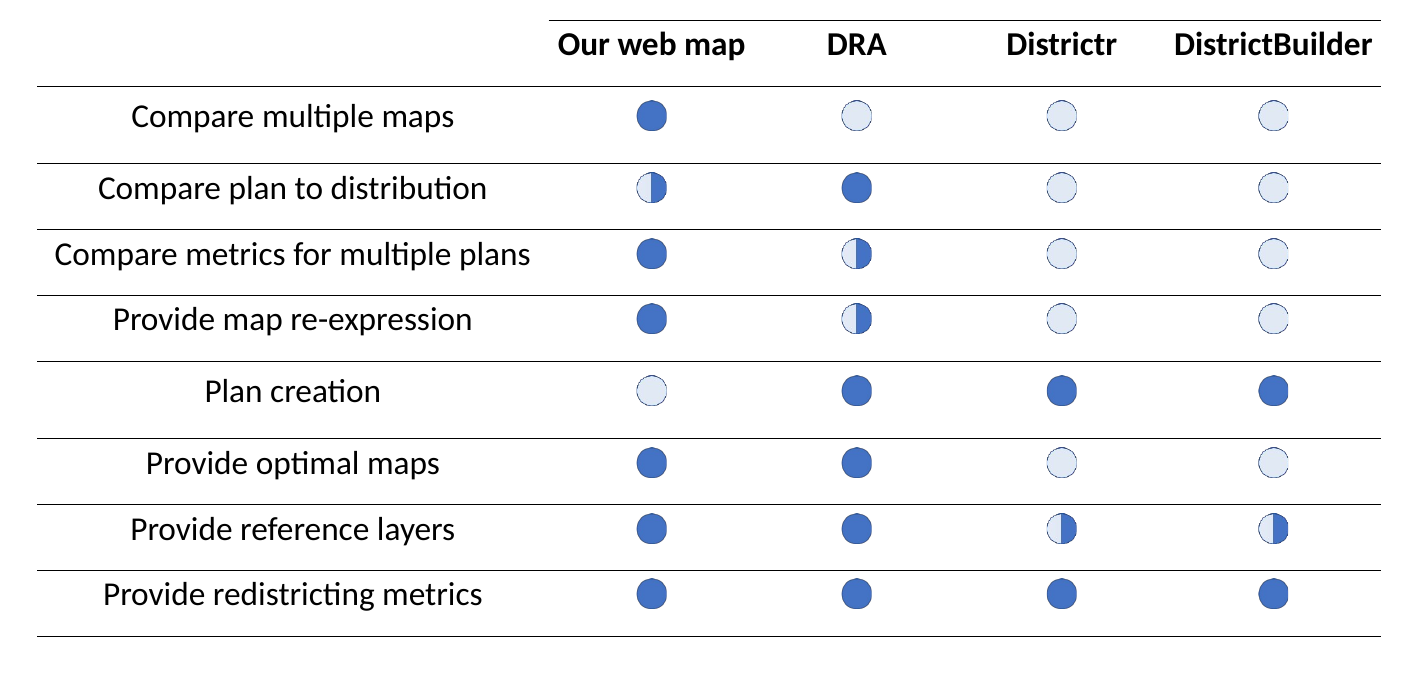}
    \caption{Comparison of our interactive web map with other redistricting websites. Dark blue circles represent full functionality, half filled circles represent partial functionality, and light blue circles represent no functionality. [two-column width]}
    \label{fig:webmap_comp}
\end{figure}

The results of the functionality comparison between our web map and other public redistricting websites can be seen in Figure \ref{fig:webmap_comp}, where dark blue circles represent full functionality, half-filled circles represent partial functionality, and light blue circles represent no functionality. This type of functionality comparison method has been widely adopted in the cartography and visual analytics communities~\citep{roth2015user,tu2023interactive}. As seen in the first column, our website and DRA provide more functionality than the other two websites. However, ours is the only web map that allows for two maps to be compared side by side, which is important for understanding the impact of plans on specific cities and regions. While our website does provide an interactive  3D cube, which has ensemble distribution values displayed in three dimensions, it does not highlight the provided maps (optimal or otherwise) in that cube. In DRA, the selected maps for presentation are shown against box plots for ensemble distributions. For displaying district-level attributes, our website is the only website that provides metrics for two maps in a parallel coordinate plot, while DRA did allow for comparison between the metrics of various maps, but without being able to view the maps themselves. DRA provided the ability to decide which attribute is expressed in the choropleth map, but ours is the only one that provides re-expression in the form of bar plots or proportional symbol maps. Notable, all of the other maps supported plan creation, while ours does not, though that was not a goal of our website as it mainly focuses on multi-criteria redistricting plan assessment.

\begin{figure}[H]
    \centering
    \includegraphics[width=1.0\linewidth]{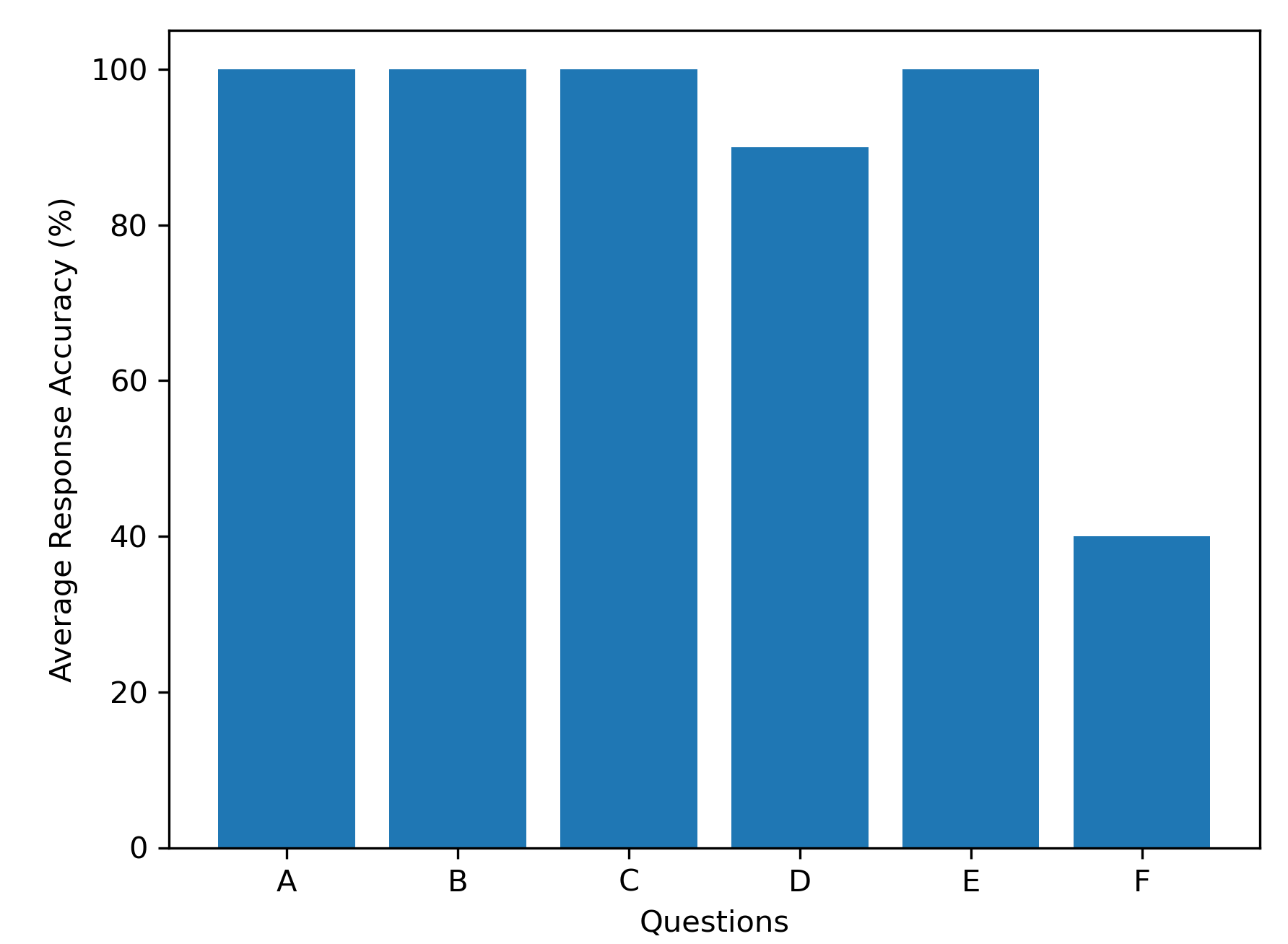}
    \caption{Average response accuracy for the usability survey. The questions are as follows: A) Add the CBG overlay to the left map pane, B) Hover over the district that contains the label ``Milwaukee" in each of the map panes, C) Change the 2022 Enacted Map panel to show the Max Compactness Map, D) Hover over the line in the parallel coordinate plot corresponding to the district with the highest inter\_flows value for the Interaction-Ratio map, E) Read the Dis\_effGap value for district containing the city of Madison in the 2022 Enacted Map, F) Hover over the point with the minimum compactness value in the interactive redistricting values 3D cube. [two-column width]}
    \label{fig:resp_acc}
\end{figure}

Figure \ref{fig:resp_acc} shows the average accuracy results (in percentage) of the usability assessment on our developed web map using the six tasked introduced in section \ref{interactivewebmap}. Response accuracy was generally very high for all map and plot reading tasks, indicating effective web design to facilitate knowledge extraction. The most difficult task for participants involved interacting with the 3D cube, suggesting that some sort of visual affordance is needed to make users aware that the 3D cube can be repositioned with mouse control, which poses a spatial cognition challenge to certain degree. Users who were able to use the cube interactively had used similar plots before in other contexts, and so assumed that the cube in the assessment was also interactive, while users who were not able to interact with the cube had no such experience. However, users were generally still able to interpret the meaning of points in the cube, as related to the three metrics used in the cube axes.

\section{Discussion}
\subsection{ReCom Sampling Modifications}
As shown in section \ref{sec:resultsampling}, the biased RST and max-IR cut methods are able to shift the sampled district plan distributions towards higher or lower IR values, as desired. We believe the reason that the algorithm was unable to sample more widely has to do with the spanning tree distribution. While this distribution yields districts that are visually compact, it is also likely restricting the space explored by our adapted algorithms. For instance, it is easy to imagine that some very non-compact districts could produce the highest IR values, if the districts were well lined-up with the underlying communities. In fact, we did set the compactness threshold even lower, to 7$\times$ the number of cut edges in the original plan, but it did not change the IR distribution over 5$\times$ the number of cut edges. This is likely because non-compact districts are rarely produced by random spanning trees, even if the compactness threshold allows for them. More generally, further research is needed to understand the spanning tree distribution used in ReCom \citep{procaccia2022compact}, and such research may shed light on how to bias the distribution of IR values sampled. 

\subsection{Relationship Between Interaction Ratio and Current Redistricting Criteria}
While a fairly monotonic relationship exists between compactness scores and IR values in this study, that may not hold for other regions. We have compared IR values to other traditional redistricting metrics (e.g., Polsby-Popper compactness and efficiency gap) so that IR can be understood in context, but, in actual use for redistricting, we would hope that the presence of actual IR-defined communities would be used as the basis for districts, rather than as a means to justify proxies like compactness. Still, the adoption of the IR for redistricting would require further study into how it relates to existing redistricting goals and criteria.

Another common redistricting constraint, which we have not addressed explicitly in this paper, is geographic contiguity (although the spanning tree method already considers the spatial connectedness of adjacent nodes/districts). Enshrined in U.S. redistricting custom is a territorial basis for districts \citep{makse2012defining}. However, there are types of communities that, while sharing common political interests, may not be geographically adjacent. Examples of this may be minority communities or virtual communities. The IR could incorporate other types of flows between districts, such as phone calls~\citep{ratti2010redrawing,gao2013discovering} or friendship link strength~\citep{bailey2018social}, to help quantify non-contiguous districts. 
More broadly, we note that spatial interaction communities may only encompass some portion of what different parties consider to be COIs, since COIs have been considered to be racial, geographic, economic, and political \citep{williams2019community}.

Finally, we note, as other have, that better capturing communities of interest may not always result in more equal seat allocation between the two major parties or more competitive districts. Indeed, the particular spatial arrangement~\citep{duchin2021political,guo2008regionalization} and partisan nature~\citep{gimpel2020conflicting} of certain types of communities may inherently make districts less competitive. 

\subsection{Interactive Web Map Assessment}
While our developed website provides functionality not present in other redistricting websites and scored highly in the usability assessment, there are limitations in the design of the assessment. Most notable, users familiar with GIS and web maps were surveyed, rather than redistricting experts with experience or interest in redistricting. Therefore, the tasks were mainly focused on usability, rather than on the sorts of tasks that people familiar with redistricting would be trying to accomplish, such as comparing the district shapes or identifying particular communities. However, our current assessment is still useful in demonstrating effective web design for the basic map-based functionalities that would facilitate insight by people exploring the map specifically for redistricting. Our future work will seek community engagement with the state-level redistricting committee to further assess our web map product.

\section{Conclusion}
In this study, we propose that spatial interaction communities be used as the basis of COIs in redistricting. To quantify spatial interaction communities, we present a new metric termed the interaction ratio (IR), which compares the global sum of intra-district population flows to the global sum of inter-district population flows for redistricting. To explore the range of IR values associated with valid district plans, we employ the ReCom algorithm on the space of graph partitions to produce an ensemble of district plans, and then calculate the IR values for all of the plans in the ensemble. As such ensembles are used frequently in legal challenges to district plans, we also propose two methods, the biased RST and min/max-IR cut methods, which allow the user to better specify the distribution of IR values that should be sampled. Using the results from these methods, we turn to a multi-criteria assessment of redistricting plans, where we show that an ensemble with higher IR values produced more proportional seat allocation outcomes for the 2018 Federal Congressional election in Wisconsin, when compared to an ensemble with lower average IR-values. We also developed a web map which allows for exploration of the redistricting results from the multi-criteria assessment produced in this research. Future research could focus on more comparisons of the IR with existing redistricting criteria, and as well as methods for changing the distribution of sampled IR values. Additionally, the relationships between IR, compactness, and proportionality in voting allocation could be explored at different spatial scales and in other regions. 

This research contributes to the political science literature on redistricting by introducing spatial interaction communities as the basis of COIs. It also contributes to computational redistricting research and practice by proposing two methods for biasing the sample distribution towards higher or lower IR values. Through the multi-criteria assessment described here and made available in the interactive web map, we provide important context for understanding how IR values relate to the values of existing redistricting metrics, which is necessary for the adoption of this new metric in redistricting and regionalization practices.

\section*{Data availability statement}
The mobility flow dataset used in this research is publicly available on GitHub:  \url{https://github.com/GeoDS/COVID19USFlows} and from SafeGraph. The other aggregated data that support the findings of this study are available from the U.S. census bureau. Due to the privacy protection policies of the data providers, the voting data used here are not publicly available.

\section*{Acknowledgments}

We would like to thank Gareth Baldrica-Franklin and Professor Robert Roth for their help and guidance in the development of the web map. We would also like to thank Professor Jin-Yi Cai for sharing his expertise on modifying the ensemble distribution in algorithmic design. This project is supported by the University of Wisconsin 2020 WARF Discovery Initiative funded project: Multidisciplinary Approach for Redistricting Knowledge. Any opinions, findings, and conclusions or recommendations expressed in this material are those of the author(s) and do not necessarily reflect the views of the funder(s).

% references APA style.
\bibliographystyle{apalike}
\bibliography{references}

\end{document}